\documentclass[aps,prb,groupedaddress,showpacs,twocolumn]{revtex4-1}

\usepackage{graphicx}
\usepackage{amsmath}
\usepackage{bbm}
\usepackage{xspace}
\usepackage{color}

\DeclareMathOperator{\diag}{diag}
\DeclareMathOperator{\Tr}{\mbox{Tr}}
\DeclareMathOperator{\tr}{\mbox{tr}}

\definecolor{orange}{RGB}{252,77,6}
\definecolor{brown}{RGB}{200,127,50}

\newcommand{\app}{App.\@\xspace}
\newcommand{\ie}{i.\thinspace{}e.\@\xspace}
\newcommand{\eg}{e.\thinspace{}g.\@\xspace}

\newcommand{\ve}[1]{{\bf #1}}
\newcommand{\mat}[1]{\mathsf{#1}}
\newcommand{\nag}{{\phantom{\dagger}}}

\newcommand{\eq}[1]{Eq.\thinspace{}(\ref{#1})}
\newcommand{\eqq}[2]{Eqs.\thinspace{}(\ref{#1}) and (\ref{#2})}

\newcommand{\eqqqqs}[4]{Eqs.\thinspace{}(\ref{#1}), (\ref{#2}), (\ref{#3}) and (\ref{#4})}

\newcommand{\fig}[1]{Fig.\thinspace{}\ref{#1}}

\newcommand{\Fig}[1]{Figure \ref{#1}}

\def\bra#1{\mathinner{\langle{#1}|}}
\def\ket#1{\mathinner{|{#1}\rangle}}
\def\braket#1{\mathinner{\langle{#1}\rangle}}

\newcommand{\ver}{\ve R}
\newcommand{\veq}{\ve q}
\newcommand{\beq}{\begin{equation}}
\newcommand{\eeq}{\end{equation}}
\newcommand{\beqn}{\begin{eqnarray}}
\newcommand{\eeqn}{\end{eqnarray}}
\newcommand{\ddiag}[1]{\diag(#1)}
\newcommand{\eeqref}[1]{Eq.\thinspace{}\eqref{#1}}
\newcommand{\ndeltaq}{\sqrt{N_c}\delta_{\veq}}
\newcommand{\tB}{\tilde B}
\newcommand{\tp}{{\bar t}}     
\newcommand{\TP}{{\bar T}}      
\newcommand{\nomat}[1]{#1} 
\newcommand{\alp}{\alpha}

\newcommand{\zero}{_{(0)}}
\newcommand{\bzh}{\text{BZ}/2}
\newcommand{\qbzh}{\ve q \in \bzh}
\usepackage{hyperref}

\begin{document}


\title{Variational cluster approach for strongly correlated lattice bosons \\in the
  superfluid phase}


\author{Michael Knap}
\email[]{michael.knap@tugraz.at}
\affiliation{Institute of Theoretical and Computational Physics, Graz University of Technology, 8010 Graz, Austria}
\author{Enrico Arrigoni}
\affiliation{Institute of Theoretical and Computational Physics, Graz University of Technology, 8010 Graz, Austria}
\author{Wolfgang von der Linden}
\affiliation{Institute of Theoretical and Computational Physics, Graz University of Technology, 8010 Graz, Austria}


\date{\today}

\begin{abstract}
We extend the variational cluster approach to deal with
strongly correlated lattice bosons in the
superfluid phase.
To this end, we reformulate the approach 
within a pseudoparticle formalism, 
whereby cluster excitations are
described by particlelike excitations.
The approximation amounts to solving a multicomponent noninteracting
bosonic system by means of a multimode Bogoliubov approximation.
A source-and-drain term is introduced in order to break $U(1)$
symmetry 
at the cluster level. 
We provide an expression for the grand potential, 
 the
single-particle normal and anomalous Green's functions,
 the condensate
density,
 and other static
quantities.
As a first nontrivial application of the method we choose the two-dimensional Bose-Hubbard model
and evaluate results in both the Mott and the superfluid phases. 
Our results show an excellent agreement with quantum Monte Carlo calculations.
\end{abstract}


\maketitle

\section{\label{sec:introduction}Introduction}

Cluster approaches have been proven to be very useful for the numerical investigation of
strongly correlated many-body systems. 
These approaches consist in
embedding finite size clusters, for which a numerical exact solution
is available,
within a lattice of infinite size. 
The embedding is done by introducing
additional 
fields to the cluster Hamiltonian, in order to 
take into account the coupling to the rest of the lattice
in some appropriate dynamical mean-field way. 
We will term these fields Weiss fields, since they
play an analogous role as in Weiss mean-field theory of ferromagnetism (\textit{cf.} Ref.~\onlinecite{ge.ko.96}).
Different cluster embedding techniques, such as cluster perturbation
theory\cite{gros_cluster_1993,snchal_cluster_2002} (CPT),
variational cluster
approach\cite{potthoff_variational_2003,potthoff_self-energy-functional_2003-1,da.ai.04} (VCA) ,
 cellular dynamical mean field theory\cite{ko.sa.01} (C-DMFT), and
dynamical
cluster approximation,\cite{he.ta.98} differ by the nature of the
Weiss fields and of the mean-field treatment which fixes their optimal value.
In the present paper, we consider VCA, which has been applied
to a large variety of fermionic~\cite{potthoff_variational_2003,da.ai.04}
and bosonic~\cite{koller_variational_2006,aichhorn_quantum_2008,knap_spectral_2010}
systems. VCA can be 
understood in a more general framework called self-energy functional
approach~\cite{potthoff_self-energy-functional_2003-1,potthoff_self-energy-functional_2003}
(SFA),
in which 
the grand potential of the physical system is expressed as the
stationary point
of a particular functional of
the self energy.
Here, we will adopt an alternative approach to VCA in which
single-particle excitations are expressed in terms of ``pseudoparticles,'' which are 
similar to Hubbard operators,~\cite{ovchinnikov_hubbard_2004} and 
external fields are ``added'' to the cluster Hamiltonian and
``subtracted perturbatively.''~\cite{zacher_evolution_2002}
We discuss in Sec.~\ref{sec:m1} the advantages of this alternative approach.

Strongly correlated lattice bosons are currently in the focus of
research due to seminal experiments on ultracold gases of
atoms.\cite{jaksch_cold_1998,greiner_quantum_2002,bloch_many-body_2008}
In these experiments quantum mechanical interference effects can be
observed
on a macroscopic scale. In particular, ultracold gases of atoms on a lattice
undergo a quantum phase transition from the Mott phase, in which particles
are localized on individual lattice sites, to the delocalized
superfluid phase, in which $U(1)$ symmetry is broken and a finite fraction of the particles forms a
Bose-Einstein condensate (BEC). 

Up to now, bosonic VCA has been formulated for the normal phase only.  
The principal aim of this paper is to extend this formalism to 
the symmetry-broken, superfluid phase. The theoretical framework, developed in the next two sections, is applicable to 
a large class of lattice boson 
systems in the Mott insulating as well as in the superfluid
phase. 
In particular, besides the widely studied Bose-Hubbard
model,~\cite{fisher_boson_1989,jaksch_cold_1998} the method can be  
straightforwardly extended to include disordered systems or 
multiple components containing, for example, fermion-boson mixtures.
The extended VCA theory can
be applied even to the $U(1)$ broken, superfluid phase of light-matter
systems, where photons are confined in coupled, nonlinear
quantum-electrodynamics
cavities.\cite{hartmann_quantum_2008,tomadin_many-body_2010}
In order to achieve the extension to the $U(1)$ broken, superfluid phase, it proves convenient to reformulate VCA in terms of 
a pseudoparticle approach, whereby single-particle excitations within
a cluster are approximately mapped 
onto 
particlelike excitations. 
We show that this approach, first applied to normal bosons, quite naturally
suggests the extension to the superfluid case. In a following
publication,~\cite{functional} we show that the results obtained
from the pseudoparticle formalism in the superfluid phase can be equivalently obtained
within an appropriate extension of the SFA taking into account condensed bosons. One of the aims of the present paper is to illustrate the advantages
of the pseudoparticle formalism, which can be
used to extend VCA to a large variety of problems with strongly correlated
lattice systems.

 The pseudoparticle formalism is
in some aspects related to the standard basis matrix operator method
developed by Haley and Erd{\"o}s in
Ref.~\onlinecite{haley_standard-basis_1972} and to the
Hubbard-operator approach, see for instance
Ref.~\onlinecite{ovchinnikov_hubbard_2004}. 
The idea is
to introduce pseudoparticle
 operators $b_{\mu}^\nag$ and
$b_{\mu}^\dag$, which connect
the ground state $\ket{\psi_0}$ with single-particle excited states
 $\ket{\psi_\nu}$ of a Hamiltonian
describing disconnected clusters in the lattice. In the VCA language the cluster Hamiltonian is termed
reference system $\hat H'$.
Of course, the
$b_{\mu}^\nag$ and
$b_{\mu}^\dag$
 do not have the properties of ordinary
single-particle creation and annihilation operators.
The crucial point is that by treating them as such, one recovers the very 
same results as obtained from CPT and from
VCA, as has been
shown for fermionic systems in Ref.~\onlinecite{zacher_evolution_2002} (see
appendix therein).
In Sec.~\ref{sec:m1} we prove the same result for the bosonic
(normal) case, which 
is somewhat more subtle, as it 
requires a
multimode Bogoliubov transformation.
In this picture, excited states
$\ket{\psi_\mu}$ are treated
as pseudoparticle excitations with
the properties
\[
\ket{\psi_\mu} = b_{\mu}^\dag \ket{\psi_0}
\quad\quad
b_{\mu}^\nag \ket{\psi_\nu}=\delta_{\mu\nu}\ket{\psi_0} \;.
\]
Within the VCA approximation, pseudoparticles are regarded as noninteracting particles.
We stress that, while this may seem a rather crude approximation, it
is equivalent to CPT and VCA. Furthermore, with appropriate extensions it
becomes equivalent to C-DMFT. 

It is straightforward to show (see Sec.~\ref{sec:m1})  that 
within this approach
 the original bosonic operators $a_i^\nag$ and $a_i^\dag$ can be expressed
as linear combinations of the pseudoparticle operators $b_\mu$.
This makes it possible to write the coupling of the cluster to the rest of the lattice, which in VCA consists of intercluster hopping terms,
 as a quadratic form
in the $b_\mu$. In combination with the fact that the cluster
Hamiltonian is by construction quadratic in these operators as well, one
finally obtains a Hamiltonian which is completely quadratic in 
the 
pseudoparticle operators, and can, thus, be solved exactly.

Our paper is organized as follows:
In Sec.~\ref{sec:m1} we first show that the
standard VCA results for the 
Green's functions and for the
grand potential $\Omega$ 
are
recovered within the pseudoparticle formalism applied to normal
bosonic systems.
In order to be able to
treat the superfluid phase we then extend the theory in
Sec.~\ref{sec:m2} by introducing 
Weiss fields in the form of ``source-and-drain'' terms which
explicitly break the
$U(1)$ symmetry of the reference Hamiltonian $\hat H'$.
The main result of
Sec.~\ref{sec:m2} is the expression for the grand potential 
$N_c \Omega$
of the physical system 
($N_c$ is the number of clusters).
Within our extension of VCA to the superfluid phase 
we obtain\cite{trace} 
\begin{align} 
  \Omega &= \Omega^\prime - \frac{1}{2N_c}\Tr\ln (-G)  + \frac{1}{2N_c}\Tr\ln
  (-G') \nonumber -\frac12 \tr h\\
  & \label{omega}
+  \frac12 \braket{A^{\dag}}  G\zero^{-1} \braket{A}
-  \frac12 \braket{A^{\dag}}' G\zero^{\prime -1} \braket{A}' \;.
\end{align}
The first three terms on the right hand side 
are essentially identical
to those which are also present in standard VCA 
expressions. Particularly,
 $\Omega'$ is the grand potential 
(per cluster)
of the reference system, and
 $G'$ and $G$ are the connected Green's functions of the
reference and of the physical system, respectively.
 However, they are
expressed in the Nambu representation, which explains the additional factor
$1/2$ and the fourth term in comparison with previous results.\cite{koller_variational_2006,aichhorn_quantum_2008,knap_spectral_2010}
The suffix $(0)$ used in the second line of \eq{omega} means that the
corresponding Green's functions are calculated for $\veq=\ve 0$ and
$\omega=0$, where $\ve q$ is the superlattice vector associated to the cluster tiling, and
$\omega$ is the Matsubara frequency.
As usual within VCA theory, the two Green's functions share the same
self-energy. 
The expectation values
$\braket{A}'$ and $\braket{A}$ are the
corresponding condensate densities, again in Nambu 
(vector)
notation.
The latter are connected 
by the relation
\beq
\label{afa}
 G\zero^{-1} \braket{A} =  F + G\zero^{\prime -1} \braket{A}'
 \;,
\eeq
where 
the vector $F$ describes the 
strength of the source-and-drain term 
 which is introduced in the
reference system in order to explicitly break $U(1)$ symmetry. The value of $F$ [see \eq{fveq}]
has to be determined from the variational principle.
Details for the notation are provided in Sec.~\ref{sec:m1}, and~\ref{sec:m2}.
In addition to the formula for the grand potential $\Omega$, we evaluate expressions for other quantities, which are useful for describing the superfluid phase. In particular, we derive expressions for the normal and anomalous Green's functions, the particle density, and the condensate density. In Sec.~\ref{sec:results}
this extended VCA theory is applied to the two-dimensional 
Bose-Hubbard (BH)
model in the superfluid phase.
Finally, we summarize and conclude our
findings in Sec.~\ref{sec:conclusion}.

\section{\label{sec:m1}Pseudoparticle approach}

In this section we reformulate CPT/VCA within the pseudoparticle
approach for bosonic systems. 
In principle, one may argue 
that
the formulation of CPT/VCA using
  pseudoparticles is complicated and in the case of the normal phase  (\ie
  Mott phase) CPT/VCA can be obtained from simpler approaches, 
as, for example, from
  Dyson's equation (see, e.g. Ref.~\onlinecite{snchal_cluster_2002}) and the SFA.\cite{koller_variational_2006,potthoff_self-energy-functional_2003-1}
The reason why we present this alternative formulation here is that 
this approach, while not as rigorous as SFA, 
provides useful hints on how to deal with more complicated 
situations, like the superfluid phase discussed 
in this work
(see Sec.~\ref{sec:m2}). 
In addition, it gives insight on other properties.
For example, in the case of normal bosons the pseudoparticle
approach is useful in order to understand the occurrence of noncausality of the
Green's function
  in cases, where the chosen reference system is not suitable to
  describe the phase of the physical system, as we point out below.
 Thus the aim of this section is to derive the principal
  theoretical framework of the pseudoparticle approach, for the normal
  phase, reproducing the known result for the grand potential
  $\Omega$, which has to be optimized. The extension to the superfluid
  phase is the subject of the next section. 

The physical system of interacting particles is described by
a grand-canonical Hamiltonian $\hat H$, which is related to the 
canonical Hamiltonian in the usual way by the additional single-particle term $-\mu \hat N$. 
The Hamiltonians, which can be treated by the extended VCA 
theory, 
generally have the form $\hat H = \hat H_t + \hat H_U$, where $\hat H_t$ consists of arbitrary one-particle terms and $\hat H_U$ of local two-particle terms.
The physical system is defined on a large or even infinite lattice with periodic boundary 
conditions. The underlying lattice is now tiled into $N_c$  
clusters 
each one containing $L$ orbitals (sites).
We split the Hamiltonian into a cluster part  $\hat H_{cl}$, which only describes processes within the various clusters, and the residual part
$\hat T$, containing the intercluster processes, which consist of single-particle terms only, so that
\beq
\label{hclt}
\hat H = \hat H_{cl} + \hat T\;.
\eeq

CPT amounts to first solving for the Hamiltonian
$\hat H_{cl}$ and then carrying out a perturbation expansion
in the intercluster Hamiltonian $\hat T$.
Of course, within CPT one is free to add 
 an arbitrary
single-particle Hamiltonian 
$-\hat \Delta$ to the cluster Hamiltonian $\hat H_{cl}$ provided 
it is then subtracted  
from $\hat T$ so that $\hat H$
remains unchanged. This defines a new cluster Hamiltonian $\hat H'$
\beq
\label{hp}
\hat H' \equiv \hat H_{cl} - \hat \Delta  
\;.
\eeq
The physical Hamiltonian $\hat{H}$, given in \eq{hclt}, can now be expressed in terms of the new cluster Hamiltonian 
\begin{equation}
\hat H = \hat H' + \hat \Delta + \hat T \equiv \hat H' + \hat\TP\,,
\label{eq:hphysfull}
\end{equation}
leading to a new ``perturbation'' $\hat\TP \equiv \hat \Delta + \hat T$.
The CPT expansion is now carried out
in this new ``perturbation''. 
While ideal exact results should not depend on $\hat \Delta$
(this occurs, for example, in the noninteracting case),
in practice results do depend on $\hat \Delta$ due to the approximate nature of the expansion. 
 The idea is to fix the parameters
$\hat \Delta$ by an optimization prescription,
 which amounts to finding the stationary point of the grand potential $\Omega$ obtained from the perturbative expansion.
The optimization prescription is put on a rigorous framework within
the SFA.~\cite{potthoff_self-energy-functional_2003}
It is straightforward to show that this procedure is equivalent to the
standard VCA prescription, whereby $\hat H'$ is the corresponding {\em reference system}.~\cite{ai.ar.05}

In the following, we consider 
$N_c$
identical disconnected
clusters, and 
denote the sites 
(orbitals) 
within a cluster by $i$.
The position of each cluster on the large, physical lattice is specified by a lattice vector $\ve R$.
Accordingly,  we denote by
$a_{i,\ve R}$ the annihilation operator for a boson on site $i$ of 
cluster $\ve R$, and similarly for creation operators
$a_{i,\ve R}^\dag$.
In order to keep a compact notation we combine the annihilation
operators of a given cluster $\ve R$ into a column vector of operators
\[
\ve{a}_\ver = (a_{1,\ve{R}},\,a_{2,\ve{R}}, \ldots
\, a_{L,\ve{R}})^T
\;,
\]
 and correspondingly, the creation
operators are row vectors
$\ve{a}^{\dag}_\ve{R} = 
(\ve{a}_\ve{R})^{\dag}
$.
Using these expressions we rewrite the intercluster Hamiltonian as
\beq
\label{hatt}
\hat T = 
\sum_{\ve{R}\,\ve{R}^\prime} \ve{a}_\ve{R}^\dagger
\nomat{t}(\ve{R}-\ve{R}^\prime) \ve{a}_\ve{R^\prime}^\nag \;,
\eeq
where $\nomat{t}(\ve{R}-\ve{R}^\prime) $ is a matrix describing the
hopping terms from cluster $\ve R^\prime$ to cluster $\ve R$, with the
property $\nomat{t}(\ve{R}-\ve{R}^\prime) =
\nomat{t}(\ve{R}'-\ve{R})^{\dag}$. Here we have assumed
translation invariance
by a cluster translation vector.
Similarly, we can express $\hat \Delta$ in terms of an intracluster
hopping matrix $h$
\[\hat \Delta = \sum_{\ve R} \ve{a}_\ve{R}^\dagger \nomat h \ \ve{a}_\ve{R} \; ,\]
such that $\hat \TP$, defined in \eq{eq:hphysfull}, can be written as \eeqref{hatt} with the replacement
\[
\nomat{t}(\ve{R}-\ve{R}^\prime) \rightarrow 
\tp(\ve{R}-\ve{R}^\prime) = \nomat t(\ve{R}-\ve{R}^\prime) +
\delta_{\ve R,\ve R'} \nomat h \;.
\]
As explained above,
the reference system consists of a sum of Hamiltonians acting
on independent clusters $\ve R$
\[
\hat H' = \sum_{\ve R} \hat H'(\ve R)\;.
\]
Again considering translation invariance, all 
$\hat H'(\ve R)$ are identical. Thus it suffices to 
determine numerically 
the ground state
$\ket{\psi_0,\ver}$, 
as well as  single particle
 or single-hole excited states
 $\ket{\psi_\mu,\ver}$ of a single cluster Hamiltonian   
$\hat H'(\ve R)$, with corresponding
 eigenenergies
$E_0^\prime$ and
$E_{\mu}^\prime$, respectively.
The key idea of the approach, to be presented here, is to
introduce pseudoparticle operators $b_{\mu,\ver}^\dag$ and
$b_{\mu,\ver}^\nag \equiv (b_{\mu,\ver}^\dag)^\dag$, 
which are defined by their matrix elements
\beq
\label{pseudob}
\bra{\psi_\mu,\ver} b_{\nu,\ve R}^\dag \ket{\psi_0,\ver} =
\delta_{\mu,\nu}\;.
\eeq
In other words, the pseudoparticle operator $b_{\mu,\ver}^\dag $
applied to  the exact many-body groundstate $\ket{\psi_0,\ver}$ of a
cluster 
creates
 the exact excited many-body state $\ket{\psi_\mu,\ver}
$. In this respect, it is of course forbidden to apply a second
pseudoparticle creation operator on the excited state. This leads to
the  
supplementary 
hard-core constraints 
\hbox{$b_{\nu,\ve R}^\dag b_{\mu,\ve R}^\dag  \ket{\psi_0,\ve R}=0$}.
To neglect this hard-core constraint and to restrict to
single-particle and single-hole excitations within each cluster is the
approximation made here. We show below that this approximation, combined
with the variational procedure discussed above, gives the same results
as
VCA. In particular, we obtain the same expression for the grand
  potential $\Omega$,
and for the Green's function.
 It should be mentioned, however,
  that within the pseudoparticle approach there is no known rigorous variational principle for $\Omega$. 
One can simply heuristically state that the ``best'' solution is the
one that 
``minimizes'' the energy, although, as we know from VCA, the
variational solution is not always a minimum.  Also for parameters,
such as the chemical potential, for
which $\Omega$ turns out to be a maximum, one can argue that 
the stationary condition is a kind of ``constraint'' fixing the
consistency of thermodynamic quantities,\cite{aichhorn_antiferromagnetic_2006} 
and the corresponding
parameter is a kind of ``Lagrange multiplier.''
Nevertheless, it is not the goal of the present paper to discuss this issue.
Here, we want simply use this ``tool'' in order to formulate an
extension of the theory to address the bosonic superfluid phase (see Sec.~\ref{sec:m2}).
 The  knowledge of the correction to the order parameter and of the grand-potential
 $\Omega$ can then guide and
  facilitate a rigorous extension of SFA to deal with the superfluid
  phase. This is the goal of a future publication.\cite{functional}  

With the help of these operators, it is straightforward to write down
a Hamiltonian
which has the same energies and eigenvectors as the reference system, 
restricted within the subspace of single-particle and single-hole
excitations from the ground state
\begin{equation}
 \hat H^\prime = N_c \ \Omega^\prime + 
\sum_{\ver} \sum_{\nu  } 
\Delta E^\prime_{\nu} \, b_{\nu,\ver}^\dagger
 \, b_{\nu,\ver}^\nag
\text{,}  
\label{eq:Href2} 
\end{equation} 
with the (positive) excitation energies $\Delta E^\prime_{\nu} \equiv
E_{\nu}^\prime - E_0^\prime$.
Since we are interested in zero temperature $T=0$, 
the grand potential of the reference system is $\Omega^\prime \equiv E_0^\prime$.

To proceed further, we need an expression for $\hat \TP$, and, thus, of the
original bosonic operators $a_{i,\ver}$, in terms of the
pseudoparticle operators. For simplicity, we drop the $\ver$ index and
concentrate on a given cluster.
Within the pseudoparticle approximation the  operators must coincide
only within the constrained subspace. 
We thus approximate each $a_i$
by an operator $\hat O_i^\nag(b_\mu^\nag,b_\mu^\dagger)$ which
shares the same matrix elements $\bra{\psi_0} \cdot \ket{\psi_0}$,
$\bra{\psi_0} \cdot \ket{\psi_\nu}$, and  $\bra{\psi_\nu} \cdot
\ket{\psi_0}$.
We express $\hat O_i$ by means of the {\em ansatz}
\begin{equation}
 \hat O^\nag_i(b_\mu^\nag,b_\mu^\dagger) = 
\sum_{\mu =1}^{n_{p}} R_{i,\mu}^\nag
{b}^\nag_{\mu} +   
\sum_{\mu =n_{p}+1}^{n_{s}}
 Z_{i,\mu}^\nag {b}^\dagger_{\mu} + \gamma_i^\nag \mathbbm{1}\;\text{,}
 \label{eq:oid}
\end{equation}
where
the first sum contains the $n_{p}$ indices associated with the 
single-particle excitations, and the second sum contains the $n_{h}$ indices for the single-hole excitations. The total number of excitations taken into account is $n_{s} = n_{p} + n_{h}$.
Here 
we have exploited
 particle-number conservation.
Next, we use this expression to evaluate the following matrix elements
\begin{subequations}
\begin{align}
 \bra{\psi_0} \hat O^\nag_i(b_\mu^\nag,b_\mu^\dagger) \ket{\psi_0} &= \gamma^\nag_i \stackrel{!}{=} \bra{\psi_0}  a_i\ket{\psi_0} \\
 \bra{\psi_\nu} \hat O^\nag_i(b_\mu^\nag,b_\mu^\dagger) \ket{\psi_0} &= Z_{i,\nu} \stackrel{!}{=} \bra{\psi_\nu}  a_i\ket{\psi_0} \\
 \bra{\psi_0} \hat O^\nag_i(b_\mu^\nag,b_\mu^\dagger) \ket{\psi_\nu} &= R_{i,\nu} \stackrel{!}{=} \bra{\psi_0}  a_i\ket{\psi_\nu} \;\text{,} 
\end{align}
\label{eq:pseudeparticlesavg}
\end{subequations}
where the coefficients $\gamma_i^\nag$ are zero so far, since the reference system
conserves the particle number. 
We now introduce
the compact notation 
\[
B
 \equiv ( b_1, \dots, b_{n_p},b_{n_p+1 }^\dag\dots
   b_{n_s}^\dag )^T
\quad\quad B^\dag = (B)^\dagger \;,
\]
\ie, the first part of the vector acts on particle states, and the
second part on hole states. 
 Notice that in this form
$B^\dagger$ ($B$) changes the number of particles by $+1$ ($-1$). 
We also introduce the $\nomat Q$ matrix (which is the same as in
Ref.~\onlinecite{knap_spectral_2010}) as
\[ 
Q_{i,\nu} \equiv \left\{ \begin{array}{ll}
           R_{i,\nu} & \text{ for $1\le \nu\le n_p$} \\
           Z_{i,\nu} & \text{ for $ n_p<\nu\le n_s$} 
      \end{array} \right. \;.
\]
The $\nomat Q$ matrix can be used to
express
the original operators $\ve a$ and $\ve a^\dagger$ in terms of $B$ operators [\textit{cf.} \eq{eq:oid}] in a compact form:
\begin{subequations}
\begin{align}
  \ve a^\nag &= \nomat Q  B \\
  \ve a^\dagger &= B^\dagger \nomat Q^\dagger \;\text{.}  
\end{align}
\label{eq:pseudeparticlesamat}
\end{subequations}
Using the compact vector notation for $B$ and $B^\dag$, the reference Hamiltonian [\eq{eq:Href2}] can be written as
\begin{align}
 \hat H^\prime &= N_c \Omega^\prime 
  + \sum_\ver B_\ver^\dagger \nomat S \nomat\Lambda B_\ver - 
N_c \Delta E'_h \;,
\label{eq:Href3}
\end{align}
where we reintroduced the $\ver$ dependence. Here we introduced the diagonal  matrices
\[
\nomat S \equiv \diag(\underbrace{1,\,\dots\,,1}_{1,\,\dots\,,n_p},
                \underbrace{-1,\,\dots\,,-1}_{n_p+1,\,\dots\,,n_s}) 
\]
and 
\[
\Lambda = \nomat S \ddiag{\Delta E_1',\,\dots\,,\Delta E'_{n_p},\,\Delta E'_{n_p+1}\,\dots\,,\Delta E_{n_s}'} \;.
\]
Notice that $\nomat S^2=1$, while  $\Lambda$
contains the poles of the  Green's function for the reference system.
The constant
\[
\Delta E'_h \equiv \sum_{\mu=n_p+1}^{n_s}\Delta E'_{\mu} =
-\tr g(\Lambda)\;,
\]
with the function
\[
g(\epsilon) \equiv \epsilon \Theta(-\epsilon) 
\]
takes into account 
that some of the boson operators have been rearranged in order to obtain  \eq{eq:Href3}.
The physical Hamiltonian introduced in \eeqref{eq:hphysfull} reads
\[
\hat H = \hat H' + \sum_{\ver,\ver'} 
\ve a^\dag_\ver  \tp(\ve{R}-\ve{R}^\prime) \ve a_{\ver'}\;.
\]
Using \eqq{eq:pseudeparticlesamat}{eq:Href3} yields a quadratic expression in
 the $B$ operators: 
\begin{align*}
\hat H &= N_c \Omega'
+N_c \tr g(\Lambda)
  + \sum_\ver B_\ver^\dagger \nomat S \Lambda B_\ver 
\nonumber \\ 
&+
  \sum_{\ver,\ver'} 
B^\dag_\ver \nomat \ Q^\dag \ \tp(\ve{R}-\ve{R}^\prime) 
 \ \nomat Q \ B_{\ver'} \;.
\end{align*}
We can now introduce a Fourier transform in the cluster vectors $\ve
R$
\begin{align}
 B_\veq &= \frac{1}{\sqrt{N_c}}  \sum_\ver
e^{i \ver \cdot \veq} B_\ver 
\nonumber \\
&=
(b_{1,\veq},\dots,b_{n_p,\veq},b_{n_p+1,-\veq}^\dag,\dots,b_{n_s,-\veq}^\dag
)^T \label{eq:fourier}
\;,
\end{align}
leading to
\beq
\label{htot}
\hat H = N_c \Omega'
+N_c \tr g(\nomat \Lambda) 
  + \sum_\veq
\hat H_\veq
\eeq
with
\begin{equation}
\hat  H_\veq \equiv B_\veq^\dag \nomat S \nomat M_\veq  B_\veq \;.
\label{eq:hq}
\end{equation}
Here, we have introduced  the matrix 
\[
\nomat M_\veq \equiv \Lambda +\nomat S \ \nomat Q^\dag  \ \tp_\veq \ \nomat Q \;,
\]
where
\[
\tp_\veq \equiv \sum_\ver e^{i\ \veq \cdot \ver}  \ \tp(\ver) 
\]
  is the Fourier transform of
$\tp(\ver-\ver')$. The non-Hermitian matrix $\nomat M_{\ve q}$ is identically defined as in Ref.~\onlinecite{knap_spectral_2010}.

Being quadratic in the $B$ operators,
$H_{\ve q}$ can be quite generally put into diagonal form by a multimode Bogoliubov transformation. To achieve this, we look for ``normal-mode'' pseudoparticles described by the vector 
$P$ with the same structure as $B$ (in the following considerations we omit the $\veq$ dependence for simplicity)
\[
P^\dagger 
 \equiv ( p_1^{s'_1}, \,\dots \,
   p_{n_s}^{s'_{n_s}} )
\quad\quad P = (P^\dagger )^\dagger \;,
\]
where 
$s'_i=\pm 1$ so that
$p_i^{+1}\equiv p_i^\dag$ is a creation and
$p_i^{-1}\equiv p_i$ is an annihilation operator. 
The new
$P$ operator shall be
connected with $B$ via 
\[
B = \nomat{V} \, P \;,
\]
where $V$ is a nonsingular but in general nonunitary matrix. From a physical viewpoint the nonsingularity of $V$ corresponds to a pseudoparticle conservation, meaning that there are as many pseudoparticles $B$ as normal-mode pseudoparticles $P$. The
transformation $V$ must satisfy two conditions. First it must be chosen such that $P$ 
has appropriate bosonic commutation relations, \ie, 
\[
[P,P^\dag]=\nomat S' 
\equiv\ddiag{s'_1,\dots,s'_{n_s}} \;.
\]
This gives
\begin{align*}
\nomat S' &{\stackrel{!}{=}}  [P,P^\dag] 
  = \nomat{V}^{-1} [B,B^\dag] (\nomat{V}^{-1})^{\dag} 
  = 
\nomat{V}^{-1} \nomat S (\nomat{V}^{-1})^{\dag} \;,
\end{align*}
which in turn yields
\begin{subequations}
\begin{align}
\label{sv}
\nomat{V} \ \nomat S' \ 
\nomat{V}^{\dag} \ \nomat S &= \nomat I \\
\ \nomat S' \ \nomat{V}^{\dag} \ \nomat S &= \nomat \nomat{V^{-1}}  \\
\  \nomat{V}^{\dag} \ \nomat S  \ \nomat \nomat{V} &= \ \nomat S'  \;.
 \label{sv_b}
\end{align}
\label{svfull}
\end{subequations}

The second requirement on $\nomat V$ is 
\beq
\label{vd}
\nomat{V}^\dagger \, \nomat S \ \nomat M \ \nomat{V} \equiv 
E \equiv \ddiag{e_1,\dots,e_{n_s}} \;,
\eeq
since  after the transformation from $B$ particles to $P$ particles the Hamiltonian in \eq{eq:hq} has to be diagonal.
Multiplying \eeqref{vd} from the left by 
$ \nomat V \nomat S'$ 
and using \eq{sv} yields the eigenvalue equation
\[ \nomat M \nomat{V} =  \nomat V \nomat D\;,\]
where $D\equiv \ddiag{d_{1},\dots,d_{n_s}} = S'E$ contains the eigenvalues of the non-Hermitian matrix 
$\nomat M$.
From \eq{eq:hp} below, where we express the Hamiltonian in terms of the normal-mode pseudoparticles, it can be seen that the diagonal elements $e_i$ correspond to the excitation energies of the physical system. Since the energy of the physical system must be bounded from below, all $e_i$ have to be positive and real, leading to 
\[
e_{i} = d_{i} s'_{i} > 0 \quad \forall i\;.
\]
It will turn out that this stability condition is the only point, where the variables $s'_{i}$ of the auxiliary operators $p_{i}^{s'_{i}}$ show up. In App.~\ref{vands} we show that, if $\nomat M$ is completely diagonalizable with real eigenvalues and linear independent eigenvectors, which is of course not generally guaranteed for a non-Hermitian matrix $\nomat M$ but necessary from the physical viewpoint, then $V$ can be constructed so that both requirements of \eqq{svfull}{vd} are fulfilled, and we can proceed with our analysis. If $M$ is not completely diagonalizable or does not have real eigenvalues, the system is unstable, and it favors a different phase, which cannot be addressed by the reference system in this form. This instability toward a different phase, such as superfluidity, has to be cured by extending the reference system by proper additional variational parameters, as discussed in Sec.~\ref{sec:m2}.

In terms of the $P$ operators we obtain for the Hamiltonian
\begin{align}
\hat H_\veq &= B_\veq^\dagger \, \nomat S \nomat{M}^\nag_\veq \,  B_\veq^\nag = P_\veq^\dagger \,
\nomat{V}_\veq^\dagger \, \nomat S \nomat{M}^\nag_\veq \, \nomat{V}_\veq^\nag \, P_\veq^\nag =
P^\dagger_\veq \nomat S' \nomat {D}^\nag_\veq \, P^\nag_\veq \nonumber\\
 &=\sum_{\nu} e^\nag_\nu 
(\, p_{\nu,\veq}^\dagger \, p_{\nu,\veq}^\nag \ \Theta(\nomat S'_{\nu,\nu})+ 
  p_{\nu,\veq}^\nag  \, p_{\nu,\veq}^\dagger   \ \Theta(-\nomat S'_{\nu,\nu}) 
)
\nonumber\\
 &= \sum_{\nu} e^\nag_\nu \, p_{\nu,\veq}^\dagger \, p_{\nu,\veq}^\nag + 
\sum_{\nu} e^\nag_\nu  \Theta(-\nomat S'_{\nu,\nu}) 
\nonumber\\
&= \sum_{\nu} e^\nag_\nu \, p_{\nu,\veq}^\dagger \, p_{\nu,\veq}^\nag - \tr  g( \nomat{D^\nag_\veq} )\;.\label{eq:hp}
\end{align} 
In the last line we have exploited the fact that
 in order for the system to be stable, \ie, the energy be
bounded from below, all $e_\nu$ must be positive. 

Inserting this expression in \eeqref{htot} yields the Hamiltonian in 
terms of diagonal normal modes.
From this result one obtains immediately the grand-canonical ground-state
energy per cluster 
\[\Omega =
 \Omega'+\tr g(\Lambda)
  - \frac{1}{N_c} \sum_\veq  
\tr  g( \nomat{D_\veq} ) \;.
\]
As discussed, $\Lambda$ and $D_\veq$ are diagonal matrices 
containing the poles of the reference Green's 
function and physical Green's function, respectively. Therefore, this expression being equivalent to Eq.~(11) in Ref.~\onlinecite{knap_spectral_2010} (see also
Refs.~\onlinecite{koller_variational_2006,sene.08u,potthoff_self-energy-functional_2003}) is equivalent to the
zero-temperature VCA grand potential.

By using the expression for the Green's function of the noninteracting
normal modes
\[
\ll p_\alpha ; p^\dag_\beta \gg = \frac{\delta_{\alpha,\beta}}{\omega -
  e_\alpha} \;,
\]
we readily obtained 
the  Green's function for the physical system
\begin{align*}
\nomat G_{\ve q}(\omega)  &\equiv \ll a_{\ve q}^\nag;\, a_{\ve q}^\dag \gg
= \nomat{Q} \ll  B_{\ve q};\, B^\dagger_{\ve q} \gg \nomat{Q}^\dagger
\nonumber \nonumber\\ 
&= 
 \nomat{Q} V_\veq \ll  P_{\ve q};\, P^\dagger_{\ve q} \gg V_\veq^\dag \nomat{Q}^\dag
\nonumber \nonumber\\ 
 &=
 \nomat{Q} V_\veq  (S'\omega  - S'D_\veq \ )^{-1} V_\veq^\dag \nomat{Q}^\dag
\nonumber \nonumber\\ 
&=
 \nomat{Q} V_\veq  (\omega - D_\veq)^{-1} S' V_\veq^\dag \nomat{Q}^\dag \;.
\end{align*}
By simple algebra this expression can be rewritten such that it is independent of the auxiliary quantities $S'$ and $V$ and thus equivalent to
Eq.~(12) in Ref.~\onlinecite{knap_spectral_2010}
\begin{align}
 \nomat G_{\ve q}(\omega)  &= \nomat{Q}  V_\veq (\omega  - V_\veq^{-1}  M_\veq  V_\veq )^{-1} V_\veq^{-1} S  \nomat{Q}^\dag 
\nonumber \\
\label{gmd}
& = \nomat{Q}   (\omega  -   M_\veq   )^{-1}  S  \nomat{Q}^\dag  \;,
\end{align}
where we have used \eqq{svfull}{vd}.

We, therefore, succeeded in proving that, for normal bosons,
 the pseudoparticle approach
yields the same Green's function and grand potential as 
VCA. For fermions, this was shown in Ref.~\onlinecite{zacher_evolution_2002}, see
appendix therein.
This result holds for $T=0$, although extension to $T>0$ is
straightforward.

\section{\label{sec:m2} Superfluid phase}

When trying to apply VCA to bosonic lattice systems in regions of the phase
diagram outside the Mott phase, one encounters instabilities which
manifest in the form of noncausal Green's functions,
\ie, in spectral functions with negative (positive) spectral weight
for positive (negative) frequencies $\omega$, or in complex poles.
Within the pseudoparticle approach 
these instabilities show up as
complex eigenvalues or negative 
diagonal elements of the matrix $E$. This kind of
instability is well known in approaches based on the bosonic Bogoliubov
approximation, such as the spin-wave approximation.

Quite generally, such an instability signals the occurrence of a phase
transition toward a new phase. 
Quite often, as in the case of the BH model studied in Sec.~\ref{sec:results},
 the new phase is the superfluid phase, which is accompanied by a Bose-Einstein condensation.
Bose-Einstein condensation is described by a finite value of the order
parameter $\langle a_\ver \rangle$. This suggests to include
a source-and-drain term in the reference system, which breaks the $U(1)$ symmetry of the reference system, leading to the ``perturbation''
$\hat \TP$ 
[see \eq{eq:hphysfull}]~\cite{notation}
\begin{equation}
 \hat \TP =  \sum_{\ve R,\ve R'} \ve{a}_\ve{R}^\dagger\, \nomat \bar t(\ve R
 -\ve R')
 \,\ve{a}_\ve{R'}^\nag + \sum_{\ve R} ( \ve{a}_\ve{R}^\dagger
 \ve{f}^\nag_\ve{R} + \ve{f}^\dagger_\ve{R} \ve{a}_\ve{R}^\nag )
 \;\text{,} 
 \label{eq:Href4}
\end{equation}
where $\ve{f}_\ve{R} \equiv (f_1,\,f_2\,\ldots\,f_L)^T$ is
a vector of size $L$ and is identical for all clusters. The index
$\ve{R}$, however, will be kept for notational reasons.

Due to these terms,
the reference system Hamiltonian does not conserve particle number anymore. Its eigenstates will thus consist of superpositions of states
with different particle numbers. 
Numerically, a
cutoff in the maximum number of boson is necessary in order to solve
 the reference system on the cluster level exactly.
We again introduce pseudoparticle operators $\ve b_{\ve R}$ connecting the ground
state with excited states. Note that we cannot distinguish between particle or hole 
states anymore. The pseudoparticles are defined by \eq{pseudob} and are connected to the original
boson operators $\ve a_{\ve R}$ by means of 
\eq{eq:oid}. Now, all matrix elements in
\eq{eq:pseudeparticlesavg} are nonzero in general. Therefore,
the two sums over $\mu$ in \eq{eq:oid} are extended to $\mu=1,\dots,n_s$,
where $n_s$ is the number of excited states considered in each cluster.

For the following considerations it is convenient to express
the boson operators within a Nambu notation. For the particle
operators we introduce
in real space
\[
 A_\ver = \left(
\begin{array}{l}
\ve a_\ver \\[0.1cm] 
\ve a_\ver^{\dagger T}
\end{array}
\right) \;,
\]
which after a Fourier transformation in the cluster vectors, see \eq{eq:fourier}, becomes
\[
 A_\veq = \left(
\begin{array}{l}
\ve a_\veq \\[0.1cm]  
\ve a_{-\veq}^{\dagger T}
\end{array}
\right) \;.
\]
For pseudoparticle operators we have in real space
\[
 B_\ver \equiv (b^\nag_{1,\ver}, \, b^\nag_{2,\ver}, \, \ldots,
 b_{n^\nag_s,\ver},\,b_{1,\ver}^\dag, \, \ldots,b_{n_s,\ver}^\dag)^T \;
\]
and in $\veq$ space
\[
 B_\veq \equiv (b^\nag_{1,\veq}, \, b^\nag_{2,\veq}, \, \ldots,
 b^\nag_{n_s,\veq},\,b_{1,-\veq}^\dag, \, \ldots,b_{n_s,-\veq}^\dag)^T \;.
\]

Similarly to Sec.~\ref{sec:m1}, we have an approximate linear
relation between the $A$ operators and the $B$ operators of the form
\[
 A_\ver = \nomat Q B_\ver + \Gamma\;\text{.}
\]
After the Fourier
transformation in the cluster vectors it reads
\beq
 A_\veq = \nomat Q B_\veq + \Gamma_\veq \;.
 \label{eq:ABrelation}
\eeq
Here,
\[
\Gamma_\veq = \ndeltaq \Gamma\;\text{,}
\]
with
\[
 \Gamma = (\gamma_1, \,\gamma_2\,\ldots \,\gamma_L, \,\gamma_1^*,
 \,\gamma_2^* \, \ldots \,\gamma_L^* \,)^T \;,
\]
and the $(2L)\times(2n_s)$ matrix
\[  \nomat Q = \left( \begin{array}{cc} \nomat R & \nomat Z \\ \nomat Z^* & \nomat R^* \end{array} \right)\;\text{.}\]
The constants $\gamma_i \equiv \bra{\psi_0} a_i \ket{\psi_0}$,
will be nonzero
as the reference system does not conserve the particle
number.

In terms of pseudoparticle operators
we can again write the reference Hamiltonian 
for a cluster $\ver$,
similarly to
 \eq{eq:Href3} as
\beq
\label{hps}
 \hat H^\prime_\ver =\Omega^\prime + \frac{1}{2} B_\ver^\dag \nomat S  \Lambda B^\nag_\ver +\frac{1}{2}\tr g(\Lambda) \;\text{.}
\eeq
Here,
the matrices $\nomat S$ and $\Lambda$ have a slightly different definition
\[
\nomat S \equiv \ddiag{\underbrace{1,\dots,1}_{1,\dots,n_s},
                \underbrace{-1,\dots,-1}_{n_s+1,\dots,2 n_s}} \;,
\]
and
\[
\Lambda = S \ \ddiag{
\Delta E^\prime_1,\,\Delta E^\prime_2\, \ldots \,\Delta E'_{n_s},\,
\Delta E^\prime_1,\,\Delta E^\prime_2\,\ldots\,\Delta E'_{n_s}} \;.
\]

To express the ``perturbation'' $\hat \TP$ of \eq{eq:Href4}, we need to
introduce a similar Nambu notation for the source-and-drain terms,
which, being $\ver$ independent, become in $\veq$ space

\beqn
&& F_\veq = \ndeltaq F 
\nonumber \\ &&
\label{fveq}
F\equiv \left(
\begin{array}{l}
\ve f \\ 
\ve f^{\dag T}
\end{array}
\right) \;.
\eeqn
After the Fourier transformation in the cluster vectors, we can rewrite
\begin{align*}
\hat \TP &= \hat T + \hat \Delta =
\sum_\veq \Bigl( 
\frac{1}{2}A_{\ve q}^\dagger\, \TP_\veq \, A_{\ve q}^\nag -
\frac{1}{2} \tr \tp_\veq\  
\\ &
+ \frac{1}{2} \left[ F_{\ve q}^\dagger\,A_{\ve q}^\nag + A_{\ve
     q}^\dagger \,F_{\ve q}^\nag \right] \Bigr) \;,
\end{align*}
where $\TP_{\ve q} = \ddiag{\tp_{\ve q}, \, \tp^{T}_{-\ve q}}$.

Replacing the $A$ operators in terms of the $B$ operators
with the help of \eq{eq:ABrelation},  
and combining \eq{hps} with the expression above for $\hat \TP$, we
finally obtain the complete Hamiltonian, defined in \eq{eq:hphysfull}, in terms of pseudoparticles
\begin{align*}
 \hat H &=N_c \Omega^\prime + \frac{N_c}{2}\tr g(\Lambda)
 + \sum_\veq  \Big\lbrace
-\frac12 \tr \tp_\veq \nonumber 
\\ 
  &+ \frac12 \Gamma_\veq^\dagger  
\TP_\veq \Gamma_\veq + 
 \frac12 B_\veq^\dag
  \underbrace{\left[ \nomat S \Lambda + \nomat Q^\dag \TP_\veq \nomat Q
    \right]}_{S \nomat M_\veq} B_\veq \nonumber
 \\ 
  &+ \frac12  \big[ {(\Gamma_\veq^\dag \TP_\veq +
    F_\veq^\dag)}
 \nomat Q B_\veq + F_\veq^\dag \Gamma_\veq + h.c. \big]\Big\rbrace\;.
\end{align*}
The expression can be further simplified by using the fact that
$F$ and $\Gamma$ 
are equal in all clusters, and thus
have only $\veq=\ve 0$ components. 
In addition we take advantage of
\beq
\sum_\veq \tr \tp_\veq = N_c \tr \tp(\ver-\ver'=0) = N_c \tr h \;,
\eeq
since $t(\ver-\ver'=0)=0$
is a pure intercluster term.
For notational convenience we introduce
\begin{equation}
\tilde F^\dag = F^\dag + \Gamma^\dag \TP_{\ve{0}}\;.
\label{eq:ftilde}
\end{equation}
This gives
\beqn
 \hat H && = N_c \Omega^\prime + \frac{N_c}{2}\tr g(\Lambda) -\frac{N_c}{2} \tr h
  + \frac{N_c}{2}\Gamma^\dagger   \TP_{\ve{0}} \Gamma 
\nonumber \\ && 
+ \frac{N_c}{2} (F^\dag \Gamma+h.c.) 
  + \frac{\sqrt{N_c}}{2} ( {\tilde   F^\dag} \nomat Q B_{\ve{0}}   + h.c. )
\nonumber \\ && 
+ \frac12 \sum_{\veq} 
 B_\veq^\dag  S\nomat M_\veq B_\veq 
\;.
\label{hwithb}
\eeqn
 
The term linear in $B$ can be eliminated by a shift
\[
\tB_\veq \equiv B_\veq + X_\veq \;,
\]
where clearly only the $\veq=\ve 0$ term of $X_{\ve q}$ is nonzero.
Considering only the $\veq=\ve 0$ part of \eq{hwithb}, which we term $Y_\ve{0}$, and plugging in the shifted operators, we obtain
\begin{align*}
Y_\ve{0}&\equiv \frac{1}{2} 
(\tB_{\ve 0}-X_{\ve{0}})^\dag  S \nomat M_{\ve{0}} (\tB_{\ve{0}}-X_{\ve{0}}) 
\\
&+ \frac{\sqrt{N_c}}{2} ( {\tilde   F^\dag} \nomat Q (\tB_{\ve{0}}-X_{\ve{0}}) + h.c. )
 \;.
\end{align*}
The linear term is eliminated by setting 
\beq
\label{x0}
X_{\ve{0}}= \sqrt{N_c} M_{\ve{0}}^{-1}S \ \nomat Q^\dag
\tilde F \;,
\eeq
yielding for the $\ve q=\ve 0$ term above
\begin{align*}
Y_\ve{0}&=\frac12 
\tB^\dag_{\ve 0} S \nomat M_{\ve{0}} \tB_{\ve 0}
+  \frac{N_c}{2} \tilde F^\dag G_{(0)} \tilde F \,,
\end{align*}
where 
\begin{equation}
G_{(0)} \equiv G_{\veq=\ve 0}(\omega=0)= \nomat -Q \ M_{\ve{0}}^{-1} S \nomat Q^\dag \;,
\label{eq:g0}
\end{equation}
which is the Green's function defined in \eqref{g1s} but evaluated for $\ve q=0$ and $\omega=0$.
In total we have
\begin{align}\label{hwithtb}
 \hat H  &= N_c\ C
+ \sum_{\qbzh}
 \tB_\veq^\dag  S \nomat M_\veq \tB_\veq 
\end{align}
with the constant terms
\begin{align*}
C &= \Omega^\prime +\frac{1}{2}\tr g(\Lambda) -\frac{1}{2} \tr h 
   + \frac{1}{2} (F^\dag \Gamma+h.c.)
\nonumber \\ 
&+\frac{1}{2}\Gamma^\dagger   \TP_{\ve{0}} \Gamma+\frac{1}{2}\tilde F^\dag G_{(0)} \tilde F  \;.
\end{align*}
In the last term of \eq{hwithtb}, we restrict the summation over half of the Brillouin zone, which we denote by $\qbzh$, and thus removed the factor $1/2$ in front of the sum. Due to Nambu representation, two summands with $+\veq$ and $-\veq$ are identical and therefore the restriction to half of the Brillouin zone is convenient. In our convention, the $\ve q=\ve 0$ term is included in the sum and retains the factor $1/2$.

\subsection{Condensate density}
Before turning to the diagonalization of the Hamiltonian in
\eq{hwithtb}, let us evaluate the condensate density. Since there are no terms linear
in $\tB$, its expectation value $\langle \tB \rangle$ vanishes.
Therefore, 
 we can immediately calculate the condensate density
\begin{align}
 \braket{A_{\ve q}} &= 
\sqrt{N_c} \, \delta_{\ve q} \, 
 \braket{A}
\nonumber \\ 
 & =
\nomat Q \braket{B_{\ve q}} + \Gamma_{\ve q} 
 = - \nomat Q X_{\ve q}+ \Gamma_{\ve q} 
\nonumber \\ &
 = \sqrt{N_c} \, \delta_{\ve q} \, [ \Gamma +
G_{(0)}
(F + \TP_{\ve 0}\Gamma) ]
 \;\text{,}
 \label{eq:sfOrder}
\end{align}
where we used \eqqqqs{fveq}{eq:ftilde}{x0}{eq:g0}.
We now exploit the fact  that
\[
\Gamma = \braket{A}'
\]
is the condensate density in the reference system. From the Dyson equation for the Green's function of the physical and the reference system we have\cite{proofg}
\[
G_\veq(\omega)^{-1} = G'(\omega)^{-1} - \TP_\veq \;.
\]
By multiplying \eqref{eq:sfOrder} with $G\zero^{-1}$ we obtain
\begin{align*}
G\zero^{-1} \braket{A} &= G\zero^{\prime-1}\braket{A}' - \TP_{\ve 0}\braket{A}' + F
+ \TP_{\ve 0}\braket{A}' 
\nonumber \\ &= G\zero^{\prime-1}\braket{A}' + F \;,
\end{align*}
which corresponds to \eq{afa}.

\subsection{Diagonalization of the Hamiltonian}
The Hamiltonian of \eq{hwithtb} 
is finally quadratic
 and 
its diagonalization proceeds in the same way as
in Sec.~\ref{sec:m1}.
Again we introduce $P$ operators 
\[ \tilde B_{\ve q} = \nomat V_{\ve q} P_{\ve q}\;,\]
and find the solution of the non-Hermitian eigenvalue equation
\[
M_\veq V^\nag_\veq = V_\veq D_\veq\;,
\]
where $V_\veq$ satisfies the relation
\[
V_\veq S' V_\veq^\dag S = I\;.
\]
The diagonal matrix $S'$, which is in principle $\veq$-dependent as well,
consists of $+1$ or $-1$ terms. It 
is chosen according to the prescription derived in App.~\ref{vands}.
The stability condition is again that the pseudoparticle eigenenergies
\[
S'D_\veq = \ddiag{e_{1\veq},\dots,e_{2n_s\veq}}
\]
are all positive.
 The physical Hamiltonian in terms of
$P$-particles now reads 
\begin{align}
 \hat H &=   \sum_{\qbzh}  P_{\ve q}^\dag S'
 \nomat{D}^\nag_{\ve q} P^\nag_{\ve q} + N_c\ C \nonumber \\
 & =  \sum_{\qbzh} \;\sum_{\nu} e^\nag_{\nu,\ve q}
p_{\nu,\veq}^\dag p^\nag_{\nu,\veq} -  
\sum_{\qbzh} \;g(\nomat
 D_{\ve q}) + N_c\ C \;\text{.}
 \label{eq:Hphys8}
\end{align}
From that we readily obtain (see \app\ref{app:x}) the grand potential per cluster 
of the physical system $\Omega$, which is the ground state expectation
value $\braket{\hat H}/N_c$ 
\begin{align}
\Omega &=
\Omega^\prime +
 \frac{1}{2}\tr g(\Lambda)- \frac{1}{N_c}
  \sum_{\qbzh} g( \nomat D_{\ve q})
 -\frac12 \tr h \nonumber\\
  &+ \frac12 \braket{A^\dag} G\zero^{-1} \braket{A} -\frac12 \braket{A^\dag}' G\zero^{\prime -1} \braket{A}' \;. \label{eq:omega}
\end{align}

By considering the fact that $\Lambda$ and $D_\veq$ contain the poles
of $G'$ and $G$, respectively, we conclude that,
in the $T\to 0$ limit\cite{trace} 
\begin{align*}
 \lim_{T\to 0} \left[ \frac{1}{2} \Tr\ln(-G')- \frac{1}{2} \Tr\ln(-G) \right] \\=
 \frac{N_c}{2}\tr g(\Lambda)- \sum_{\qbzh} g(
   \nomat D_{\ve q}) \;.
\end{align*}
Thus, \eq{eq:omega} is equivalent to \eq{omega} in the introduction in the $T=0$ limit. An extension to $T>0$ is straight forward.

The  {\em connected} Green's function now contains anomalous contributions,
but formally is obtained as in \eq{gmd},

\begin{align}
 \nomat G_{\ve q}(\,\omega)  &\equiv \ll A_{\ve q}^\nag;\, A_{\ve q}^\dag \gg_c
= \nomat{Q} \ll \tilde B_{\ve q};\, \tilde B^\dagger_{\ve q} \gg \nomat{Q}^\dagger
\nonumber \\ 
&= 
 \nomat{Q} V_\veq \ll  P_{\ve q};\, P^\dagger_{\ve q} \gg V_\veq^\dag \nomat{Q}^\dag
\nonumber \\ 
 &=
 \nomat{Q} V_\veq  ( S'\omega - S'D_\veq )^{-1} V_\veq^\dag \nomat{Q}^\dag
\nonumber \\ 
&=
 \nomat{Q} V_\veq  (\omega - D_\veq)^{-1}  V_\veq^{-1} S
 \nomat{Q}^\dag 
\nonumber \\ &=
\label{g1s}
 \nomat{Q}  (\omega - M_\veq)^{-1} S
 \nomat{Q}^\dag 
\;,
\end{align}
where we have neglected the shifts $\Gamma$ and $X_{\ve{0}}$ since they only
contribute to disconnected parts.
Notice that \eq{g1s} is a $2L\times 2 L$ matrix in Nambu and
cluster-site space. The $\veq$ vectors above refer to the reduced Brillouin zone
originating from the cluster tiling, therefore $G$ is expressed in a
mixed representation.
In translation-invariant systems, the Green's function is expected to be 
diagonal in the wave vectors $\ve k$ of the full Brillouin zone.
This symmetry is notoriously broken in cluster methods such as VCA or C-DMFT.
In order to obtain 
a $\ve k$-diagonal
$2\times 2$ Nambu Green's function
$ G(\ve k,\,\omega)$
 we need 
to apply a periodization prescription.\cite{snchal_spectral_2000}
This gives
\[ \nomat {G}(\ve k,\,\omega) = \ve v_{\ve k}^\dag 
         G^\nag_{\ve k}(\omega)
\ve v_{\ve
   k}^\nag\;\text{,} 
\]
where
\beq
\label{vek}
 \ve v_{\ve k}^\dag \equiv 
\frac{1}{\sqrt{L}}\left(
\begin{array}{cccccc}
  e^{-i\,\ve{k}\,\ve{r}_1}&\ldots & e^{-i\,\ve{k}\,\ve{r}_{L}}& 0                        & \ldots & 0 \\
  0                       &\ldots & 0                           & e^{-i\,\ve{k}\,\ve{r}_1} & \ldots & e^{-i\,\ve{k}\,\ve{r}_{L}} 
\end{array}
\right)
\;\text{,}
\eeq
and ${\ve r_i}$ is the position of site $i$ within the cluster.

A nontrivial test for VCA is the noninteracting limit, for which
this approximation becomes exact.
In Appendix~\ref{app:a} we
carry out this check for the noninteracting BH model, \ie, we set $U=0$, and
 for a reference
system consisting of single-site clusters. In this test case the grand
potential $\Omega$ of the physical system can be evaluated
analytically both using the VCA prescription as well as directly from the Hamiltonian of noninteracting lattice bosons.

\subsection{Particle density and momentum distribution}
The total particle density is defined as
\[
 n= \frac{1}{N} \sum_{\ve q} \sum_{i} \langle a_{i,\ve q}^\dag a_{i,\ve q}^\nag\rangle \;\text{,}
\]
where $N=N_c\, L$ is the total number of lattice sites present in the physical system. The particle density can be easily expressed in Nambu formalism
\begin{align}
 n&=\frac{1}{2N}\sum_{\ve q} \sum_{i} ( \langle a_{i,\ve q}^\dag
 a_{i,\ve q}^\nag\rangle + \langle a_{i,-\ve q}^\nag a_{i,-\ve
   q}^\dag\rangle) - \frac{1}{2} 
\nonumber 
\\  &=
-\frac{1}{2} + \frac{1}{2N}\sum_{\ve q} \langle A_{\ve q}^\dag
  A_{\ve q}^\nag\rangle
\nonumber 
\\&= -\frac12 
 +\frac{1}{2N}\sum_{\ve q} 
( 
  \langle P_\veq^\dag V_\veq^\dag Q^{\dag} Q V^\nag_\veq P_\veq^\nag\rangle  +
      \braket{A_\veq}^\dag \braket{A_\veq} )
\nonumber 
\\&= 
\label{eq:nsf}
-\frac12 
 + \frac{1}{2N}\sum_{\ve q}  
\tr [ \Theta(-D_\veq^\nag)  V_\veq^\dag Q^{\dag}\ Q V^\nag_\veq ]
+  \frac{1}{2L}  \braket{A^\dag} \braket{A}\;,
\end{align}
where the last term describes the contribution from the condensate,
which can be deduced from
 \eq{eq:sfOrder}. 
The term with the sum over $\veq$ can be rewritten to obtain the known form of the particle density\cite{knap_spectral_2010}
\begin{align*}
n&= -\frac12 - \frac{1}{2N}\sum_{\ve q} 
\tr  [\Theta(-D_\veq^\nag) S' V_\veq^\dag Q^{\dag} Q V_\veq^\nag ] +  \frac{1}{2L}  \braket{A^\dag} \braket{A}
\\&=
 -\frac12- \frac{1}{2N}\sum_{\ve q} 
 \tr [\Theta(-D_\veq^\nag)  V_\veq^{-1} S Q^{\dag} Q V_\veq^\nag ] +  \frac{1}{2L}  \braket{A^\dag} \braket{A}
\;.
\end{align*}

The momentum distribution $n(\ve k)$ can be extracted
by the Fourier transform within the cluster leading to
\begin{align*}
 n(\ve k) &=
-\frac1{2N} 
+  \frac{\delta_{\ve k}}{2L}  \braket{A^\dag} \braket{A}
\\&
 + \frac{1}{2N}
\tr [
\ve v_{\ve k}^{\dag} Q V_{\ve k}^\nag  \Theta(-D_{\ve k})  V_{\ve k}^\dag
Q^{\dag}  \ve v_{\ve k}^\nag
]  \;,
\end{align*}
where $\ve v_{\ve k}^\dag$ is given by \eq{vek}.

\section{\label{sec:results}Application to the Bose-Hubbard model}
\begin{figure}
        \centering
        \includegraphics[width=0.4\textwidth]{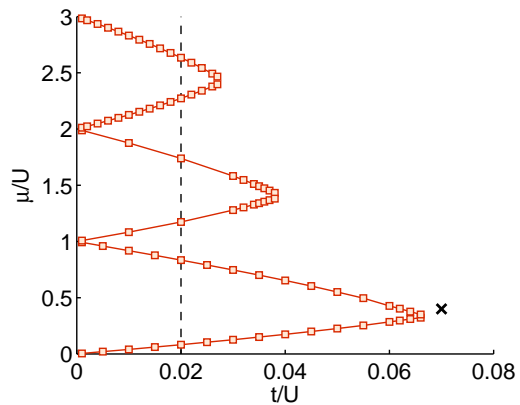}
        \caption{(Color online) Phase boundary for the first three Mott lobes corresponding to filling $n=1$, $2$ and $3$. The data for the first two Mott lobes have been published in Ref.~\onlinecite{knap_spectral_2010}. Static quantities are evaluated along the dashed line, \ie, for $t/U=0.02$ and $\mu/U$ ranging from $0$ to $3$, whereas, the dynamic single-particle spectral function is evaluated at $t=0.07$ and $\mu=0.4$, see mark $\mat x$.}
        \label{fig:pd}
\end{figure}

In this section, we present the first nontrivial application of the extended VCA theory to the two-dimensional BH
   (BH) model and compare the results with unbiased quantum Monte Carlo
  (QMC) calculations. The BH
  Hamiltonian,\cite{fisher_boson_1989,jaksch_cold_1998} which
  describes strongly correlated lattice bosons, reads 
\[ \hat{H}=-t \sum_{\left\langle i,\,j \right\rangle} a_i^\dag \, a_j^\nag
+ \frac{U}{2} \sum_i \hat{n}_i\left(\hat{n}_i-1 \right) - \mu \sum_i \hat{n}_i \; \mbox{,}
\]
where $a_i^\dag$ ($a_i^\nag$) creates (destroys) a bosonic particle
and $\hat{n}_i = a_i^\dag\,a_i^\nag$ counts the number of particles at
lattice site $i$. The parameter $t$ is the hopping strength, which
originates from
the overlap of the localized wave functions
belonging to lattice sites $i$ and $j$, respectively. The first sum 
(indicated by angle brackets) is
restricted to  ordered pairs of nearest neighbor sites. The repulsive on-site interaction is
termed $U$, and $\mu$ is the chemical potential, which controls the
particle number. For increasing ratio $t/U$ the system undergoes a
quantum phase transition from the Mott to the superfluid phase. 
We evaluate static quantities, such as the particle
density $n$ and the condensate density $n_c$ as well as the dynamic
single-particle spectral function $A(\ve k,\,\omega)$. The phase
boundary of the first three Mott lobes as obtained in VCA is shown in \fig{fig:pd}.  
The data for the first two lobes have been published in
Ref.~\onlinecite{knap_spectral_2010}. Static quantities are evaluated
for constant hopping strength $t/U=0.02$ and distinct values of the
chemical potential $\mu/U$ ranging from $0$ to $3$, scanning through various Mott lobes separated by the superfluid phase; see the dashed line in
\fig{fig:pd}. 
The single-particle spectral function is evaluated for the parameter set marked by $\mat x$ in \fig{fig:pd},
which is located in the superfluid phase close to the tip of the
first Mott lobe.
For the
numerical evaluation we used the chemical potential $\mu^\prime$ and
the strength of the source-and-drain coupling term $F$ of the reference system as
variational parameters. If not stated differently, the reference
system consists of a cluster of size $L=2\times2$. 

\begin{figure}
        \centering
        \includegraphics[width=0.4\textwidth]{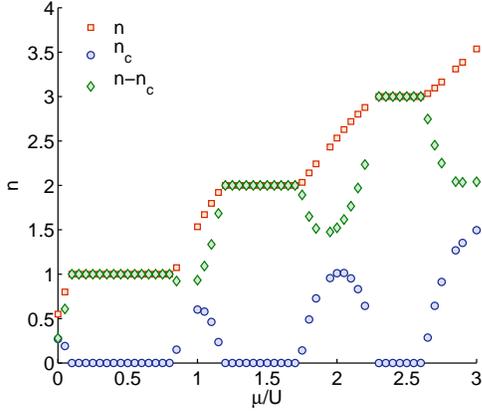}
        \caption{(Color online) Total particle density $n$, condensate density $n_{c}$, and density of the particles which are not condensed $n-n_{c}$ evaluated along the dashed line shown in \fig{fig:pd}.  }
        \label{fig:n2x2}
\end{figure}

\begin{figure}
        \centering
        \includegraphics[width=0.4\textwidth]{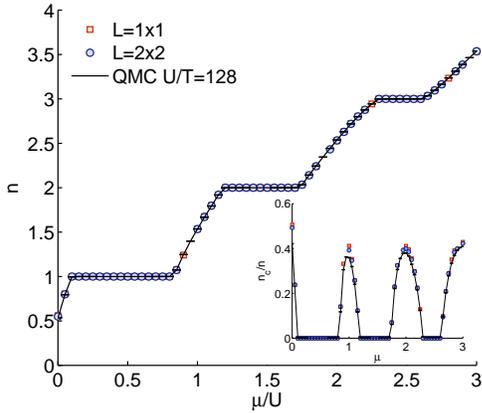}
        \caption{(Color online) Comparison of the total particle density $n$ evaluated by means of VCA and QMC 
for parameters along the dashed line in \fig{fig:pd} ($t/U=0.02$). 
The inset compares VCA and QMC results for the condensate fraction
$n_c/n$. VCA results are obtained for reference systems of size
$L=1\times1$ and $L=2\times2$ and essentially infinitely large
physical systems. QMC results are obtained for physical systems of
size $32\times32$ 
inverse
temperature $U/T=128$. }
        \label{fig:nComp}
\end{figure}

The total particle density $n$ evaluated using \eq{eq:nsf} is shown in
\fig{fig:n2x2} along with the condensate density 
$n_{c}= \braket{A^\dag} \braket{A}/2L$,
 and the density of the particles which are not condensed $n-n_{c}$. 
From \fig{fig:nComp} it can be observed that the particle density $n$ evaluated for reference systems of size $L=1\times1$ and of size $L=2\times2$ are almost identical. The same holds for the condensate fraction $n_c/n$, which is shown in the inset of \fig{fig:nComp}.
In the same figure, we also compare our results with
QMC calculations. The densities obtained from the two methods
show an excellent agreement.
The QMC data have been obtained for a system of
size $32\times 32$ and temperature $U/T=128$ using the ALPS
library\cite{albuquerque_alps_2007} and the ALPS
applications.\cite{ALPS_DIRLOOP}

\begin{figure}
        \centering
        \includegraphics[width=0.4\textwidth]{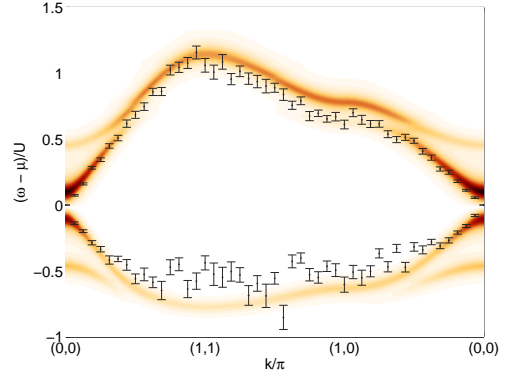}
        \caption{(Color online) Single-particle spectral function
          $A(\ve k,\,\omega)$ evaluated at $t/U=0.07$ and
          $\mu/U=0.4$. The colored density plot corresponds to VCA
          results and the dots with errorbars to latest QMC results of
          Ref.~\onlinecite{pippan_2010}.  } 
        \label{fig:sf}
\end{figure}

\begin{figure}
        \centering
        \includegraphics[width=0.4\textwidth]{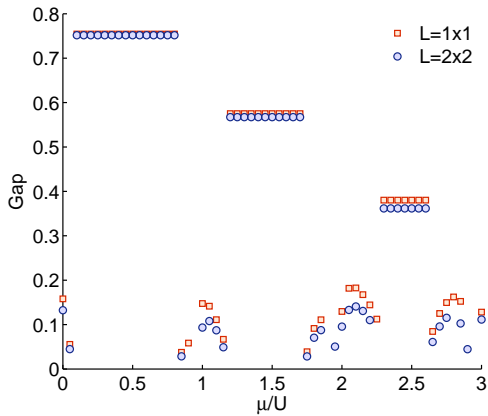}
        \caption{(Color online) Gap of the inner modes present in the single-particle spectral function measured at $\ve k = \ve 0$ and evaluated along the dashed line shown in \fig{fig:pd} for reference systems of size $L=1\times1$ and $L=2\times2$, respectively. }
        \label{fig:gap}
\end{figure}

The single-particle spectral function $A(\ve k, \,\omega)$ evaluated for the parameter set, marked by $\mat x$ in \fig{fig:pd}, \ie, in the superfluid phase close to the tip of
the first Mott lobe, is depicted in \fig{fig:sf}.
The colored density plot corresponds to VCA results and the dots with
errorbars to latest QMC results of Ref.~\onlinecite{pippan_2010}. 
The VCA spectral function $A(\ve k, \,\omega)$
consists of four bands, which is in agreement with results obtained by
means of a variational mean field
calculation,\cite{huber_dynamical_2007} a strong coupling
approach,\cite{sengupta_mott-insulator-to-superfluid_2005} and random
phase approximation (RPA)
calculations.\cite{ohashi_itinerant-localized_2006,
  menotti_spectral_2008}  
The advantage of VCA in comparison to the above mentioned approaches is that 
the results can be systematically improved by increasing the cluster size of 
the reference system.
For each wave vector $\ve k$ the weight is concentrated in one of the two bands present at positive and negative energy, respectively. 
We observe that
the outer two modes exhibit a wide gap at $\ve k = \ve 0$, which is
approximately of size $U$. The inner two, low-energy modes are also
gapped at $\ve k = \ve 0$. However, the gap is tiny, and away from $\ve k = \ve 0$ the spectrum quickly develops a linear behavior, which is 
in agreement  with the expected dispersion of  Goldstone modes.
The failure in obtaining a gapless  long-wavelength excitation is a common problem of conserving approximations, \ie, of approximations for which macroscopic conservation
laws are fulfilled.
Similar aspects occur in dynamical mean-field theory
calculations of two-component ultracold atoms as
well.\cite{hubener_magnetic_2009} In VCA there exists the additional
possibility to systematically improve the obtained results by increasing the cluster
size $L$ of the reference system. \Fig{fig:gap} compares the $\ve k =
\ve 0$ gap of the inner modes for reference systems of size
$L=1\times1$ and $L=2\times2$. 
The gap is evaluated along the dashed line shown in \fig{fig:pd}. 
The first observation is that
the gap present in the condensed  phase is almost an
order of magnitude smaller than the gap  in the Mott phase. 
It vanishes at the Mott-to-superfluid transition and, most importantly, shrinks with increasing cluster size $L$. This behavior signals
convergence toward the correct result.  

In \fig{fig:sf} we also compare our VCA results for the single-particle
spectral function to latest QMC results obtained in
Ref.~\onlinecite{pippan_2010}. In this figure QMC results are
indicated by dots with errorbars, which quantify the peak position of
the spectral weight.
Overall, we find good agreement in the low-energy spectrum. Only very close to $\ve k=\ve 0$ the two results differ slightly and the QMC dispersion possesses the  correct gapless behavior. 
The QMC spectral function, exhibits only two instead of four bands. 
This is,
however, not surprising since for the considered parameter set and at a
specific wave vector $\ve k$ the weight of one positive (negative) energy band dominates
drastically over the other one located at positive (negative) energy. Thus the four bands are extremely
difficult to resolve by means of the maximum entropy method, which has
been used to infer the spectra from QMC data; see
Ref.~\onlinecite{pippan_2010} for details concerning the QMC
results. This reference also contains a comparison between VCA data
and QMC data for the spectral function evaluated in the Mott phase, where the results obtained from the two approaches coincide very
well for all $\ve k$ values. 

We also evaluated the particle density $n$ 
for the parameter set in the superfluid phase used in \fig{fig:sf} and compared it to the QMC results.
VCA yields $n=1.0321$ in excellent agreement with the QMC result $n^{\text{QMC}}=1.03068(2)$ obtained at $U/T=128$ 
for a system of size $32\times32$.

In the following we provide some additional remarks on the fact that the single-particle spectral function, obtained within our approach, is gapped 
 in the long wavelength limit,
\ie, close to $\ve k = \ve 0$, for modes which ought to be identified
with the Goldstone modes. This issue is a commonly known problem of conserving
approximations.\cite{hohenberg_microscopic_1965} 
One condition for an approximation to be conserving
is, for example, to be $\Phi$-derivable and self-consistent (see
Ref.~\onlinecite{ba.ka.61,baym.62,luttinger_ground-state_1960,de_dominicis_stationary_1964-1,de_dominicis_stationary_1964}
for details). VCA is  $\Phi$-derivable but not self-consistent:
  The self-energy is the derivative of a functional of the Green's
  function, but the latter is not the Green's function obtained from
  Dyson's equation.
Thus VCA is not completely
conserving.
However, many conservation laws are fulfilled at the stationary point
of the self-energy functional, depending on which variational
parameters are taken into account (see Ref.~\onlinecite{or.ba.07} for a
more detailed discussion).

To obtain a gapless spectrum, a system of condensed bosons has to
fulfill an \textit{independent} condition, which is the 
Hugenholtz-Pines theorem.\cite{hugenholtz_ground-state_1959,de_dominicis_stationary_1964-1,kita_self-consistent_2009}
There are only very few systematic approximation schemes
which satisfy both conditions simultaneously.
One notable exception occurs for
interacting  bosons composed of paired fermions. 
In this case, a consistent and gapless approximation can be developed
provided the theory is  expressed in terms of the constituent fermions.~\cite{st.pi.05} 
In a different work\cite{yu.kl.06} it was suggested to include an
additional Lagrange multiplier
in the form of a chemical potential, in order to explicitly
enforce the Hugenholtz-Pines
condition. Unfortunately, the
Hugenholtz-Pines theorem is not fulfilled in VCA, and thus the
low-energy modes of the single-particle spectral function are gapped
in the long wavelength limit. Yet, the gap present in the VCA single-particle spectral function is small, and the
spectrum quickly develops a linear
behavior reminiscent of the gapless and linear Goldstone
modes. Furthermore, in VCA there exists the possibility to systematically improve the results by increasing the cluster size of the reference system.

It is also interesting to mention that the related strong coupling approximation RPA, which yields a gapless
sectrum, yet is not conserving\cite{ohashi_itinerant-localized_2006,
  menotti_spectral_2008} can be obtained within certain limits of the
extended VCA formalism. Specifically, the limits to consider are (i) to use
clusters of size $L=1\times 1$, (ii) not to use the
chemical potential $\mu$ as variational parameter and (iii) to
determine the source-and-drain coupling strength $F$ self-consistently within
a mean-field approach, whereby
  intercluster hopping terms
$ a_i^\dag \, a_j $ are replaced with their mean-field value
$ \braket{a_i^\dag} \, a_j + a_i^\dag \, \braket{a_j} $ in the
reference Hamiltonian. This leads to
the selfconsistency condition 
$
 F = z\,t\,\braket{A}\;\text{,}
$
where $\braket{A}$ is given by \eq{eq:sfOrder} and
$z$ is the coordination number of the lattice.
Our formalism provides a natural way to improve on RPA 
in a gapless, yet nonconserving, way by simply increasing the cluster
size $L$
and fixing $F$ using the mean-field condition discussed above.
However, it has to be emphasized that VCA yields much better results than RPA, even if RPA is extended to clusters of size $L$. Specifically, the particle density, the condensate density and the location of the phase boundary\cite{knap_spectral_2010,knap_benchmarkingvariational_2010} can be determined much more accurately by means of VCA only because we allowed for a variation in the chemical potential $\mu'$, \ie, allowed for macroscopic conservation laws to be fullfilled.

\section{\label{sec:conclusion}Conclusions}
In the present paper, we introduced a pseudoparticle formalism 
for interacting bosonic systems,
and showed that the results of the variational cluster approach can be derived
within this formalism. 
We used it 
to extend
the variational cluster approach to the superfluid
phase of strongly correlated lattice bosons. 
We derived 
expressions for the
grand potential and for other quantities, which are necessary to
investigate the superfluid properties.
 Our results suggest that the pseudoparticle formalism is 
a quite versatile approach,
which can be  applied to a large variety of other problems.

As a first nontrivial application of the extended version of the variational cluster approach we choose the
two-dimensional Bose-Hubbard model and evaluated static quantities
such as the total particle density and the condensate density, as well
as the dynamic single-particle spectral function. We compared the
single-particle spectral function with recent Quantum Monte-Carlo
results\cite{pippan_2010} and found good agreement between the
two approaches. It has to be pointed out that our extended variational cluster
approach, while fulfilling many conservation laws, does not fulfill the Hugenholtz-Pines
theorem. From this fact follows that the low-energy excitations of the
spectrum have a small but nonzero gap in the long wavelength limit. 
This is a common aspect, which is already present in 
theories of the dilute Bose gas.\cite{hohenberg_microscopic_1965,griffin_conserving_1996,shi_finite-temperature_1998}  However, 
for wavevectors away from
 $\ve k=\ve 0$ the spectra obtained within this approach quite soon exhibit 
a correct linear behavior and
 agree very well with the Quantum Monte-Carlo results. 
 Moreover, the gap shrinks with increasing cluster size, corroborating that the variational cluster approach becomes exact in the infinite cluster limit.
 Due to the
fact this approach fulfills several conservation laws, the particle density, the condensate density as well as the phase boundary\cite{knap_spectral_2010,knap_benchmarkingvariational_2010} delimiting the Mott from the superfluid phase can be evaluated very accurately. In the present paper we demonstrated, that our variational cluster approach results for the densities evaluated in both, the  Mott and the superfluid phase, match perfectly with Quantum Monte-Carlo results.

\begin{acknowledgments}
We made use of the ALPS library and the ALPS applications.\cite{albuquerque_alps_2007, ALPS_DIRLOOP}
We acknowledge financial support from the Austrian Science Fund (FWF)
under the doctoral program ``Numerical Simulations in Technical
Sciences'' Grant No. W1208-N18 (M.K.) and under Project No. P18551-N16
(E.A.).
\end{acknowledgments}

\appendix

\section{\label{vands} Procedure to construct  $V$ and $S'$ }

Here, we outline how the two conditions on $V$ given in \eqq{svfull}{vd} can be achieved and how $S'$ can be constructed.
We start out from the eigenvalue equation for the non-Hermitian matrix $M$
\begin{align*}
M V &= V D\;.
\end{align*}
As already argued in Sec.~\ref{sec:m1}, from the physical viewpoint we can only proceed if the eigenvector-matrix $V$ is nonsingular and 
if all eigenvalues are real, as the system would otherwise be unstable.
Hence we can express the Hermitian diagonal matrix of eigenvalues as 
\begin{align*}
D &= V^{-1} M V\;.
\end{align*}
The first condition of \eq{svfull} requires that the Hermitian matrix
\begin{align*}
X & \equiv V^{\dag} S V
\end{align*}
be diagonal with diagonal elements $X_{ii}=\pm 1$. Multiplying the two Hermitian matrices and exploiting the Hermiticity of $SM$ results in
\begin{align*}
X D &= V^{\dag} S  M V = (X D )^{\dag} = D X\;;\quad \Rightarrow[X,D] = 0\;.
\end{align*}
Commuting Hermitian matrices have a common set of orthonormal eigenvectors. 
The matrix $D$ is already  diagonal. Hence for indices belonging to nondegenerate eigenvalues, $X$ is also diagonal. 
Within the set of indices belonging to a degenerate eigenvalue, the corresponding Hermitian 
submatrix of $X$ can be diagonalized by a unitary transformation $U$. In the following we term the diagonalized matrix as $X'$. 
The diagonalization also results in a new matrix $\overline{V} = V U$ of eigenvectors. We still have 
$ \overline{V}^{-1} M \overline{V}  = D$, but now 
\begin{align}
\label{eq:xxx1}
 \overline{V} ^{\dag} S \overline{V} &=X' = \diag(x'_{1},\ldots,x'_{L})\\
\overline{V}^{\dag} S  M \overline{V} &\equiv E' =  
X'  D = \diag(x'_{1} d_{1},\ldots,x'_{n_{s}}d_{n_{s}})\;. \nonumber
\end{align}
For the condition  \eq{svfull} we still need to ensure that $x'_{\alpha}=\pm 1$.
Provided  no  $x'_{\alpha}$ vanishes, which we will show below, this can easily be achieved by a suitable normalization of the column vector of $\overline{V}\to \tilde{V} = \overline{V} Z$, with $Z$ being a diagonal matrix, defined as $Z_{\alpha\alpha} \equiv 1/\sqrt{|x'_{\alpha}|}$. We eventually have
\begin{align}
 \tilde{V}^{-1} M \tilde{V} & = D = \diag(d_{1},\ldots,d_{L}) \nonumber \\
 \tilde{V} ^{\dag} S \tilde{V} &= Z^{\dag} X' Z= S' =\diag(s'_{1},\ldots,s'_{L})\nonumber \\
\tilde{V}^{\dag} S  M \tilde{V} &\equiv   E = \diag(e_{1},\ldots,e_{n_{s}})\;. \nonumber
\end{align}

We are merely left with the proof that 
\begin{align}\label{eq:proof}
x'_{\alp}&={\ve {\bar v}}^{\alp\dag}\nomat S {\ve {\bar v}}^{\alp} 
 \ne 0\;,
\end{align}
where  ${\ve {\bar v}}^{\alp}$  stands for the  $\alp$th  column of $\overline V$.
 To this end we assume 
 {\em ad absurdum}~that ${\ve {\bar v}}^{\alp\dag}\nomat S {\ve {\bar v}}^{\alp} = 0$. In this case,
  ${\ve {\bar v}}^{\alp}$ would belong to the
$(n_s-1)$-dimensional space ${\cal S}_\alp$ 
orthogonal to the vector $S {\ve {\bar v}}^{\alp}$. According to
 \eq{eq:xxx1} the vectors
${\ve {\bar v}}^{1},\,\ldots \,{\ve {\bar v}}^{\alpha-1},\,{\ve {\bar v}}^{\alpha+1},\,\ldots\,{\ve {\bar v}}^{n_s}$ also belong to ${\cal S}_\alp$ and they are   linear independent.
Thus they span ${\cal S}_\alp$. Due to the fact that \textit{all} vectors ${\ve {\bar v}}^{1},\,\ldots \,{\ve {\bar v}}^{n_s}$ are linear independent, ${\ve {\bar v}}^{\alpha}$ cannot belong to ${\cal S}_\alp$, which proves \eq{eq:proof}.

\section{\label{app:x} Grand potential}
In this appendix we derive \eq{eq:omega}. Starting out from \eq{eq:Hphys8} we get
\begin{align}
 \Omega &= C - \frac{1}{N_c} \sum_{\qbzh} g( \nomat D_{\ve q}) 
\nonumber \\ 
&= \Omega^\prime + \frac{1}{2}\tr g(\Lambda)- \frac{1}{N_c} \sum_{\qbzh} g( \nomat D_{\ve q})  + \frac{1}{2} (F^\dag \Gamma+h.c.)  
\nonumber \\ 
\label{eq:omsf}
 &-\frac12 \tr h+ \frac{1}{2}\Gamma^\dagger   \TP_{\ve{0}} \Gamma
+\frac{1}{2}\tilde F^\dag G_{(0)} \tilde F\;.
\end{align}
We now evaluate the quantity
\begin{align*}
W & \equiv \braket{A^\dag} G\zero^{-1} \braket{A} - \braket{A^\dag}'
  G\zero'^{-1} \braket{A}' 
 \\ &=
\Gamma^\dag G\zero^{-1} \Gamma + 
\Gamma^\dag \tilde F + \tilde F^\dag \Gamma +
\\ &\quad
\tilde F^\dag G\zero \tilde F -  
\Gamma^\dag (G\zero^{-1} + \TP_\ve{0}) \Gamma 
\\ &  = 
(\Gamma^\dag ( F+ \TP_\ve{0} \Gamma) + h.c.) +
\tilde F^\dag G\zero \tilde F -  
\Gamma^\dag  \TP_\ve{0} \Gamma 
\\ &  = 
(\Gamma^\dag  F + h.c. ) + \Gamma^\dag  \TP_\ve{0} \Gamma +
\tilde F^\dag G\zero \tilde F  \;.
\end{align*}
Comparison with \eqref{eq:omsf} gives
\begin{align*}
\Omega &=
\Omega^\prime +
 \frac{1}{2}\tr g(\Lambda)- \frac{1}{N_c}
  \sum_{\qbzh} g( \nomat D_{\ve q})
+\frac12 W -\frac12 \tr h \;,
\end{align*}
which is the expression for the grand potential stated in \eq{eq:omega}.

\section{\label{app:a} Zero-interaction limit}
The zero-interaction limit turns out to be a nontrivial check for VCA.
For $U=0$, the BH model can be solved analytically as it reduces to 
\[
\hat H = -t \sum_{\langle i,\,j\rangle} a_i^\dag \, a_j^\nag - \mu \sum_i \hat n_i \;\text{.}
\]
The chemical potential $\mu$ has to be smaller than $-2t$ in order to prevent infinitely many particles in the ground state. Taking this into account, the grand potential at zero temperature is $\Omega=0$. In the zero-interaction limit VCA/CPT yields exact results. Thus the pseudoparticle formalism can be checked by applying this limit. For reference systems $\hat H^\prime$ which consist of a single site the calculations can be done analytically. Under these considerations the Hamiltonian $\hat H^\prime$ reads
\[
\hat{H}^\prime=-\mu^\prime \, a^\dagger \, a^\nag - (  a^\dagger\,f + f^*\, a)\;\text{.}
\]
It can be solved by introducing shifted operators $\tilde a \equiv a + x$ and by ``completing the square''
\begin{align*}
 \hat{H}^\prime &=  -\mu^\prime \, a^\dagger \, a - (a^\dagger\,f + f^*\, a)\\
  &\stackrel{!}{=} \alpha \,\tilde{a}^\dagger \, \tilde a^\nag + c = \alpha (a^\dagger + x^*) (a + x) + c\\
  &= \alpha \, a^\dagger\, a + \alpha ( a^\dagger x + x^* a) + \alpha \, |x|^2 +c\;\text{.}
\end{align*}
Comparison reveals
\begin{align*}
 \alpha &= -\mu^\prime \\
 x &= -f/\alpha = f/\mu^\prime \\
 c &= - \alpha \, |x|^2 = |f|^2/\mu^\prime\,\text{.}
\end{align*}
The Hamiltonian $\hat{H}^\prime$, rewritten by means of the shifted operators, is given by
\[ 
\hat{H}^\prime = -\mu^\prime \,\tilde{a}^\dagger \, \tilde a + |f|^2/\mu^\prime\;\text{.} 
\]
As discussed before we choose $\mu' < 0$. The eigenenergies obtained form the Schr\"{o}dinger equation are 
\[ 
\hat{H}^\prime \ket{\tilde \nu} = (-\mu^\prime \,\tilde \nu + |f|^2/\mu^\prime)\ket{\tilde \nu} = E^\prime_\nu\ket{\tilde \nu}\;\text{.} 
\]
For negative chemical potential $\mu'$ the ground state is
$\ket{\psi_0}=\ket{\tilde 0}$ and its energy
$E^\prime_0=|f|^2/\mu^\prime$. The eigenstates of $\hat{H}^\prime$ are
number states, therefore the shifted creation and annihilation operators act
on them in the usual way 
\begin{align*}
 \tilde a^\nag \ket{\tilde\nu} &= \sqrt{\tilde \nu} \ket{\tilde \nu -1} \\
 \tilde a^\dagger \ket{\tilde\nu} &= \sqrt{\tilde \nu+1} \ket{\tilde \nu +1}\;\text{.}
\end{align*}

To evaluate the $\nomat{Q}$ matrices we apply the original operators $a$ on the eigenstates of $\hat{H}^\prime$
\[
a \ket{\tilde \nu} = (\tilde a - f/\mu') \ket{\tilde \nu} = \sqrt{\tilde{\nu}}\ket{\tilde \nu-1} - f/\mu' \ket{\tilde \nu} \;\text{.}
\]
With that we obtain
\begin{align*}
 \bra{\tilde 0}a\ket{\tilde \nu} &= 1 \\
 \bra{\tilde \nu}a\ket{\tilde 0} &= 0 \\
 \bra{\tilde 0}a\ket{\tilde 0} &= -f/\mu'\;.
\end{align*}
Writing down the expressions in matrix form yields
\[\nomat Q = \left( \begin{array}{cc}
 1 & 0 \\ 0 & 1 
\end{array}
 \right)  = \mathbbm{1} \quad
  \Gamma = -1/\mu' \left( \begin{array}{c}
 f \\ f^*
\end{array}
 \right) \quad \nomat S =\left( \begin{array}{cc}
 1 & 0 \\ 0 & -1 
\end{array}
 \right)\]
\[ \nomat\Lambda =\nomat S \left( \begin{array}{cc}
 E^\prime_1-E^\prime_0 & 0 \\ 0 & E^\prime_1-E^\prime_0  
\end{array}
 \right) = \left( \begin{array}{cc}
 -\mu' & 0 \\ 0 & \mu'  
\end{array}
 \right)\;\text{.}\]
Using the expressions above and the relation $A = \nomat Q B + \Gamma$ we obtain for the pseudoparticle operators
\[ B = \nomat Q^{-1} (A-\Gamma) = \tilde A\;\text{.}\]
Next, we evaluate the grand potential from \eq{eq:omsf}, where we obtain 
\begin{align*}
 \Omega &=
   \underbrace{\Omega^\prime +\frac{1}{2}\tr g(\Lambda)}_{\text{A}} - \underbrace{\frac{1}{N_c} \sum_{\qbzh} g( \nomat D_{\ve q})}_{\text{B}}  +  \underbrace{\frac{1}{2} (F^\dag \Gamma+h.c.)}_{\text{C}}  \\
&-\underbrace{\frac12 \tr h}_{\text{D}}+  \underbrace{\frac{1}{2}\Gamma^\dagger   \TP_{\ve{0}} \Gamma}_{\text{E}}
 -\underbrace{\frac{1}{2}\tilde F^\dag \nomat Q \ M_{\ve{0}}^{-1} \nomat S \nomat Q^\dag  \tilde F}_{\text{F}}
\end{align*}
by employing \eq{eq:g0}. We calculate parts A--F of $\Omega$ separately
{\allowdisplaybreaks
\begin{align*}
 \text{A:  }\quad&\Omega^\prime +\frac{1}{2}\tr g(\Lambda) = |f|^2/\mu' + \mu'/2\\
 \text{B:  }\quad&\frac{1}{N_c} \sum_{\qbzh} g( \nomat D_{\ve q}) = \mu/2 \\
 \text{C:  }\quad&\frac{1}{2} (F^\dag \Gamma+h.c.)  = -2\,|f|^2/\mu'\\
 \text{D:  }\quad&\frac12 \tr h  = (\mu'-\mu)/2\\
 \text{E:  }\quad& \frac{1}{2}\Gamma^\dagger   \TP_{\ve{0}} \Gamma = |f|^2 (\mu'-\mu-2t)/{\mu'}^2\\
 \text{F:  }\quad& \frac12 \tilde{F}^\dagger \nomat{Q} \nomat{M}^{-1}_{\ve 0} \nomat S \nomat{Q}^\dagger \tilde{F} = -|f|^2\,(\mu + 2t)/\mu'^2\;\text{.}
\end{align*}
}
In order to evaluate part B we need the matrix $\nomat M_\ve{q}$, which is given by
\[
\nomat M_\ve{q} = \nomat \Lambda +\nomat S \nomat Q^\dagger \,\nomat \TP_{\ve q} \, \nomat Q = \left( \begin{array}{cc}-\mu-2\,t \cos \ve{q} & 0 \\0 & \mu + 2\,t \cos \ve{q}\end{array}\right)\;\text{,}
\]
where we used that 
\[
 \TP_{\ve q} = \left( \begin{array}{cc} \tp_{\ve q} & 0 \\0 & \tp^{T}_{-\ve q} \end{array} \right)
\]
and $\tp_\veq=\tp^{T}_{-\ve q}=\mu'-\mu-2t \cos {\ve q}$.
Since $\nomat M_\ve{q}$ is already diagonal we can readily evaluate part B as sum over the negative eigenvalues, which is $\mu + 2\,t \cos \ve{q}$, since $\mu<-2t$. When summing over half of the $\ve q$ values the second term of the eigenvalue containing $\cos \ve q$ is zero. For the calculation of part F
we need the inverse of $\nomat M_{\ve{0}}$, which is simply
\[ \nomat M_{\ve{0}}^{-1} = \left( \begin{array}{cc} -\frac{1}{\mu+2 t} & 0 \\ 0 &  \frac{1}{\mu+2 t}\end{array}  \right)\;,\]
and $\tilde F$, which reads
\begin{align*}
\tilde F &= F + \nomat \TP_{\ve{0}}\,\Gamma = (\mu+2t)/\mu'\left( \begin{array}{c} f \\ f^* \end{array}  \right)\;.  
\end{align*}
Collecting all terms yields the grand potential $ \Omega=0$, which is identical to the result obtained from the direct calculation.


\begin{thebibliography}{52}
\bibitem [{not()}]{notation}%
  \BibitemOpen
  \href@noop {} {}\bibinfo {note} {In order to limit the complexity of the
  notation, in Sec.~\ref{sec:m2} we often use symbols already defined in
  Sec.~\ref{sec:m1}. In some cases they have exactly the same definition, in
  other cases they may have a similar meaning but a slightly different
  definition. In the latter case, the symbol is redefined in
  Sec.~\ref{sec:m2}.}\BibitemShut {Stop}%
\end{thebibliography}
%

\end{document}